\documentclass[12pt]{spieman} 
\usepackage{amsmath,amsfonts,amssymb}
\usepackage{graphicx}
\usepackage{setspace}
\usepackage{tocloft}
\usepackage{textcomp}
\usepackage{physics}
\usepackage{comment}
\usepackage[export]{adjustbox}
\usepackage{lineno}
\usepackage{caption}
\usepackage{subcaption}
\usepackage{xcolor}
\usepackage[normalem]{ulem}

\title{Segment-level thermal sensitivity analysis for exo-Earth coronagraphy with segmented space telescopes}

\author[a,$\ast$]{Ananya Sahoo} 
\author[b]{Laurent Pueyo} 
\author[c]{Iva Laginja} 
\author[b]{Bryony F. Nickson} 
\author[d] {Leonid Pogorelyuk}
\author[e]{Laura E. Coyle} 
\author[b]{R\'emi Soummer} 
\author[f]{Matthew East}
\affil[a]{University of Massachusetts, 600 Suffolk Street, Lowell, MA 01854, USA}
\affil[b]{Space Telescope Science Institute, 3700 San Martin Drive, Baltimore, USA}
\affil[c]{LIRA, Observatoire de Paris, Universit\'{e} PSL, Sorbonne Universit\'{e}, Universit\'{e} Paris Cit\'{e}, CY Cergy Paris Universit\'{e}, 92190 Meudon, France}
\affil[d]{Rensselaer Polytechnic Institute, 110 8th St, Troy, NY 12180, USA}
\affil[e]{BAE Systems, Space \& Mission Systems Inc., 1600 Commerce St,  Boulder, USA}
\affil[f]{L3Harris Technologies, Inc., 332 Initiative Dr, Rochester, USA}

\cftpagenumbersoff{figure}
\cftpagenumbersoff{table}
\usepackage{xcolor}

\begin{document} 
\maketitle

\captionsetup[figure]{name=Figure}

\begin{abstract}Direct imaging and characterization of Earth-like planets require ultra-stable wavefronts to achieve a starlight suppression level of 100 parts per trillion (ppt) in the coronagraphic dark region of the host star. Thermal drifts in the telescope may affect the wavefront stability. In this context, we present a segment-level thermal stability error budget for segmented space telescopes crucial to Earth-like planet detection and specify requirements for an ultra-stable telescope. Our study utilizes multiple segmented primary mirror architectures, each with their respective apodizer solution developed by the Segmented Coronagraph Design \& Analysis research team at the Space Telescope Science Institute, tailored to an off-axis ${\sim}6$~m-aperture space telescope design. Using a detailed finite element model provided by L3Harris Technologies, we relate the temperature gradient at the location of the primary mirror to wavefront variations on each segment. We allocate both static and dynamic thermal tolerances for each segment using the Pair-based Analytical model for Segmented Telescope Imaging from Space sensitivity approach, and a batch-estimation algorithm respectively. Our analysis shows a non-uniform tolerance allocation across all segments of the primary mirror, as a result of apodization, and generally, with an increase in segment numbers, the tolerances for outer segments become less stringent. We observe trade-offs between segment size, tolerance relaxation and optimal wavefront sensing time to achieve a desired dark-hole contrast. 
\end{abstract}

\keywords{segmented telescope; thermal stability requirements; high-contrast imaging; wavefront error budget}

{\noindent \footnotesize\textbf{*}Address all correspondence to Ananya Sahoo,  \linkable{ananya$\_$sahoo@uml.edu}}

\begin{spacing}{2} 

\section{Introduction}
\label{sect:intro}  
Imaging and characterizing Earth-like planets is one of the key science objectives for the Habitable Worlds Observatory (HWO) as identified by the NASA 2020 Astrophysics Decadal Survey\cite{2020decadalsurvey}. Earth-like planets, or exo-Earths, are extremely faint bodies (i.e., ${\sim}10^{10}$ times fainter than their host star), typically located at a close angular separation (${\sim}0.1$ arcsecond). To directly image them we need telescopes with high resolving power and large collecting area, thus requiring a large diameter for the primary mirror (PM)\cite{stark2014maximizing}, and a coronagraph to occult the host star light. In addition, active wavefront sensing and control (WFS\&C)  will be necessary to suppress the residual diffraction effects and create a stable dark hole (DH) surrounding the host star. Previous studies \cite{stark2014maximizing, stark2019exoearth} assume a DH contrast floor of $10^{-10}$ per spatial resolution element and a contrast stability of at least $10^{-11}$ in the visible band at small inner working angles, typically between $\sim 1-5\lambda/D$,  of the coronagraphic point spread function (PSF) to maximize the yield of exo-Earths\cite{mamajek2024nasa, nemati2020method, nemati2023analytical, stark2024paths}. For this, the telescope system will need to operate in an ultra-stable regime, requiring wavefront stabilization at the level of $10^{-12}$~m\cite{lyon2012space,coyle2019large, nemati2017effects}. Achieving such a wavefront stability requires maturing ultra-stable structures in the picometer regime and a system-level error-budget to flow mission-level performance requirements down to sub-systems and components\cite{coyle2024achieved}. Imperfect optics or misalignment in the system will result in a non-ideal wavefront for the coronagraph, which gives rise to speckles in the DH and degrade the contrast. WFS\&C using one or more deformable mirrors (DM)s is often used to sense and improve the contrast. Set to launch in September 2026, the Nancy Grace Roman Space Telescope (RST) will use two DMs in space to create DHs for direct imaging of exoplanets\cite{cady2025high}. A typical RST observing cycle will include (a) pointing the telescope at a bright reference star to find the optimal DM shapes using WFS\&C, (b) slewing to the science target with those constant DM shapes, (c) rolling the telescope on the science target a few times to enable the removal of star light from the planet light during post-processing stages, (d) and then slewing back to the reference star to re-create the optimal DH, which may have degraded due to the telescope's slew or orientation\cite{krist2023end}, and changes in the thermal environment. Note that, continuous WFS\&C \cite{redmond2022implementation, redmond2024exoplanet}, can compensate for most static instrumental effects or drifts and maximize the performance of the coronagraph onboard the telescope. Therefore, while specifying requirements for an ultra-stable telescope, it is important to consider both the coronagraph and the wavefront sensing and control-loop as an integrated system, where the requirements can be relaxed using adjustable optics\cite{pueyo2022coronagraphic}. Future large space telescopes, such as the HWO may include segmented mirrors to optimize mission scale and mass and to adapt to possible launch vehicles. The presence of segment gaps and phasing errors between segments introduces complex diffraction patterns, thus making high-contrast imaging very challenging \cite{nemati2017effects, pueyo2014high}. Moreover, thermal distortions due to stellar flux or internal onboard-instruments, gravitational forces, electrostatic forces, radiation and other observatory dynamics will affect the optical stability. Each of these error sources needs to be quantitatively addressed in order to build a picometer-level wavefront error (WFE) budget\cite{coyle2024achieved}.

In this article, we focus on the impact of surface deformations occurring on segments due to thermal heating of their mounting pads and set thermal stability requirements for individual segments. This work is based on a sub-analysis conducted within the broader ULTRA (Ultra-stable Large Telescope Research and Analysis) Program\cite{coyle2019large, coyle2022achieved, coyle2024achieved, Coyle2019UltraStableTelescopeResearch}, which is a system engineering study involving industry, academia and nonprofits partners to mature critical technologies in support of the HWO mission concept. The study identifies thermo-elastic effects and dynamic perturbations associated with PM back-plane support structure as sources of wavefront variation affecting picometer-level wavefront stability, and thereby limiting exo-Earth detection. Here we present a methodology to compute segment-level thermal stability requirements for an arbitrarily segmented space telescope architecture, using WFS\&C to mitigate risks associated with wavefront instability. We use a finite element model (FEM) provided by L3Harris Technologies to relate temperature gradients at location of the PM to wavefront variations. These models are localized surface deformations on a segment, caused by 1~mK temperature gradient surrounding the segment. Here, we examine hexagonal segment shapes; however, this method is also applicable to other segment shapes like pie wedge and keystone, which are currently being considered in the Exploratory Analytic Case (EAC) study phases \cite{feinberg2024habitable}. In \S~\ref{sec:fem}, we describe the FEMs for a hexagonal segment and their impact on wavefront stability. In \S~\ref{sec:pastis}, we apply the Pair-based Analytical model for Segmented Telescope Imaging from Space (PASTIS) framework\cite{laginja2021analytical} to constrain static wavefront error and thermal deviation for each segment necessary to achieve a DH contrast stability of $10^{-11}$. In particular, we build a contrast-based sensitivity matrix which relates spatially averaged DH contrast to small aberrations at the PM. This is done by converting the FEM's thermally-induced surface figure distortion into WFE and then propagating that through a diffractive model of a segmented PM with a coronagraph. We invert the sensitivity matrix to obtain static tolerances for each segment required for exo-Earth imaging. We then compare the requirements obtained across different Segmented Coronagraph Design \& Analysis (SCDA) telescope architectures with their optimized Apodized Pupil Lyot Coronagraph (APLC) \cite{soummer2004apodized, n2016apodized} designs. In a realistic observing scenario, the observatory's thermal conditions may drift over time due to telescope orientation/slew, which induces dynamic WFE. Subsequently, new WFE temporal tolerance values that take continuous close-loop wavefront sensing and control into account are required. In \S~\ref{sec:temporal}, we translate the static requirements obtained in \S~\ref{sec:pastis} into a time domain. For this, we implement a batch-estimation algorithm introduced by Pogorelyuk et al. (2021) \cite{pogorelyuk2021information} to relate the static variance semi-analytically to closed-loop variance of the segments’ thermo-mechanical mode. The close-loop variances and limiting contrast for each segment-level finite-element thermal mode, and for each of the PM architectures, comprising varying segment sizes and numbers, are summarized in \S~\ref{sec:conclude}.
 
\section{Segment-level response to 1~mK temperature change}\label{sec:fem}
\begin{figure}[!ht]
\centering
\includegraphics[width=\linewidth]{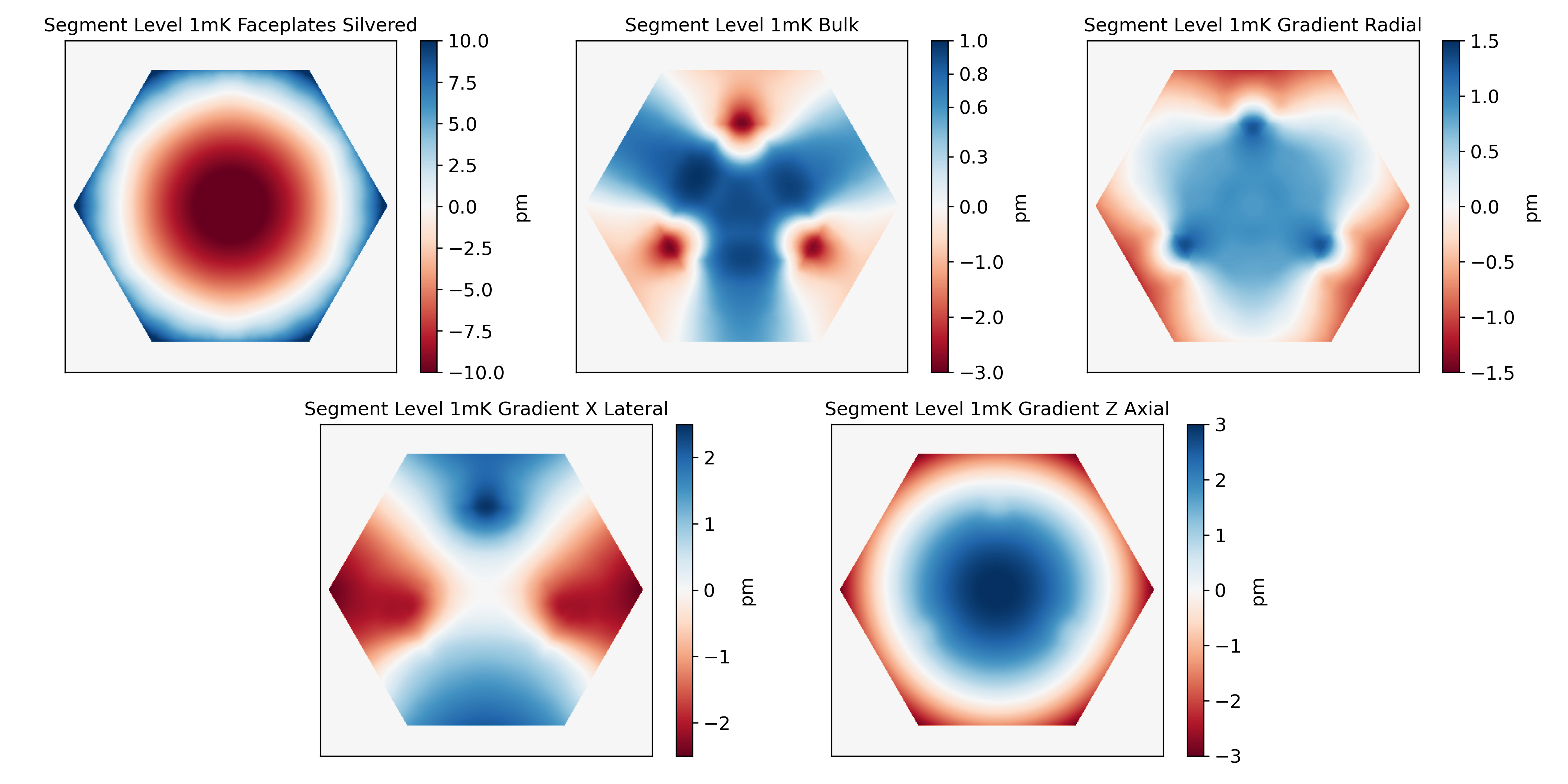}
 \caption{Segment-level surface deformations due to 1~mK temperature change along different axial directions. These deformations are of the order of picometer, and may ultimately limit the coronagraphic performance required for exo-Earth detection. We use these finite element models to estimate optical sensitivities due to thermal drifts. \textit{Courtesy: BAE Systems Inc. and L3Harris Technologies.}} \label{fig:thermal_basis}
\end{figure}

In space, small thermal and mechanical disturbances in the observatory can significantly distort the instrument wavefront giving rise to quasi-static speckles in the image plane, and thus degrade the raw DH contrast. Thermo-elastic effects will induce deformations of optical components or misalignments in the optical train, creating wavefront instabilities. Often, Structural, Thermal and Optical (STOP) analysis is used to predict the optical performance of a telescope under a predefined set of thermal loads. It involves generating a FEM which maps the thermally induced deformation to the geometry of the telescope structure. In this paper, we confine our study to the optical performance related to the thermo-elastic effects associated with back-plane support structure of a segmented primary mirror and use five FEMs provided by L3Harris Technologies to assess the coronagraphic performance. The primary mirror segment assembly (PMSA) model considered in the ULTRA study, primarily consists of an ULE\textsuperscript{\textregistered} mirror substrate, adhesive invar bond pads, invar struts and kinematic flexures for constraining rigid body motion. The adhesive bond pads being in direct contact with the mirror substrate induces surface deformations when a temperature change of 1~mK is applied along different axial directions of a segment\cite{coyle2021technology, Coyle2019UltraStableTelescopeResearch}.  The sensitivity of these surface deformations are of the order of pm mK~$^{-1}$  (see Fig.~\ref{fig:thermal_basis}), and may ultimately limit the performance of the coronagraph. 

High-order WFS\&C techniques such as Pairwise Probing with Electric Field Conjugation (EFC)\cite{give2011pair, thomas2010laboratory} and Speckle Nulling\cite{borde2006high} can be used to reduce static speckles inside the DH. Fast low-order aberrations can be compensated using a fast-steering mirror or a low-order Zernike Wavefront Sensor\cite{pourcelot2023low}. However, over a long integration time the optimal solutions for DM state given by these high-order control algorithms will be inadequate to compensate the slow WFE drifts\cite{redmond2021dark, pueyo2022coronagraphic}. The correction in such cases is limited by the number of photons, DM actuators profile: dynamic range and counts, detector non-linearity, model inaccuracies, temporal instabilities such as vibrational and thermal drifts, among many other factors. The level of wavefront correction achievable with the current state-of-the-art DMs tends to be typically within few nanometers \cite{soummer2024high, mennesson2024current}. Recent laboratory demonstrations of WFS\&C using DMs with a ZWFS has shown that low- and mid- spatial frequency aberrations can be sensed and controlled down to few picometers-level \cite{ruane2020wavefront, ruane2020microelectromechanical}. The effect of these picometer-level surface deformations, likely to contribute to the slow high-spatial frequency aberrations, might not be fully corrected. Therefore, it becomes necessary to study their impacts on wavefront stability, and accommodate these sensitivities while building a picometer-level error-budget for exo-Earth imaging. As these design-dependent thermal models are more likely to occur than localized Zernike polynomials, we directly use them as the basis set for our analysis and study the coronagraphic impact of these five kinds of segment-level surface deformations on the DH contrast. 

\section{Using the PASTIS approach to set segment-level static thermal requirements}\label{sec:pastis}
\subsection{Method}
\begin{figure}[!ht]
\centering
\includegraphics[width=0.7\linewidth]{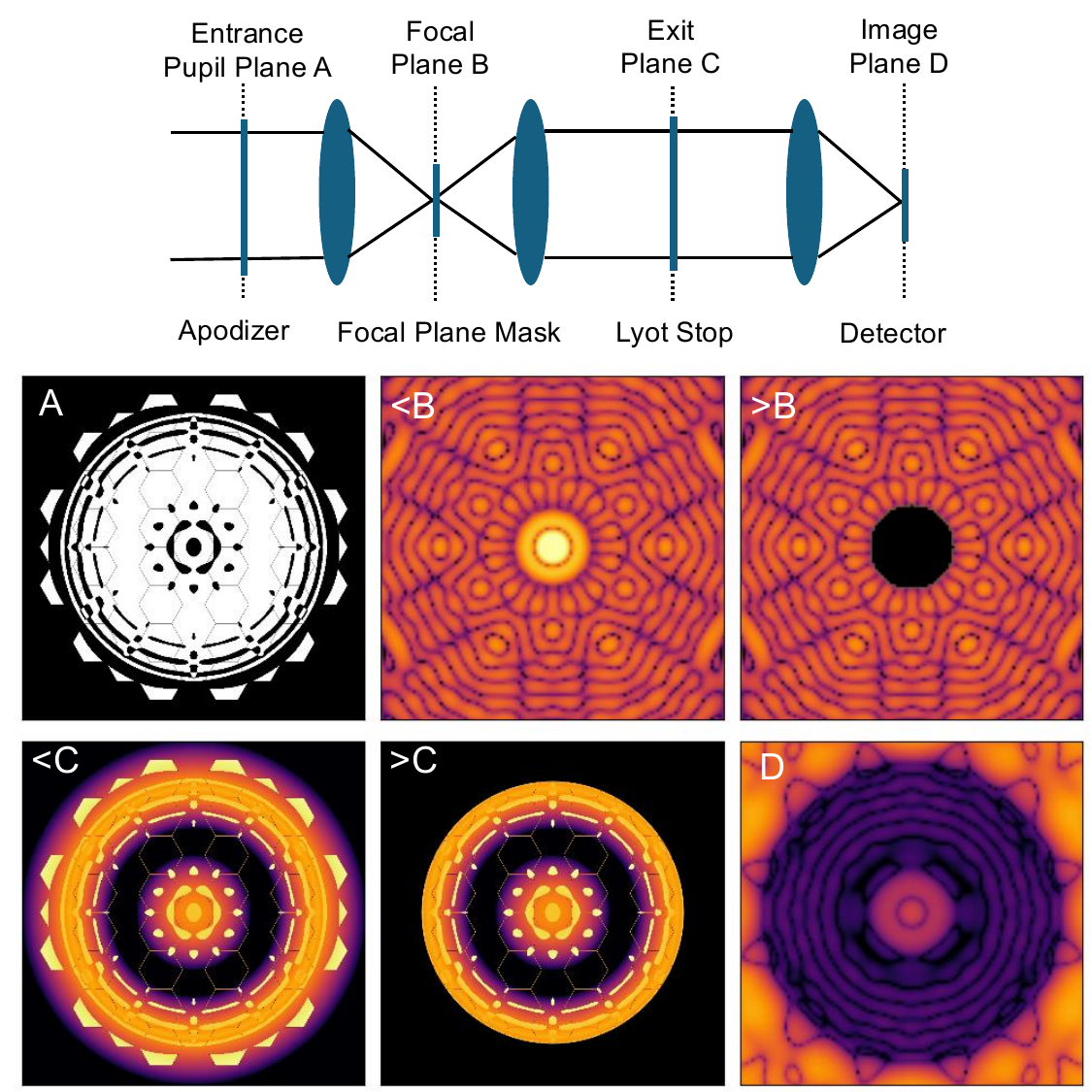}
 \caption{Diffractive optical model of a segmented PM architecture with an APLC coronagraph. A binary mask is placed in the pupil plane A to modulate the amplitude of incident star light, followed by a circular focal plane mask (FPM) in the plane B to block the on-axis light, and a circular Lyot stop in the conjugate pupil plane C to remove residual diffracted light by the FPM. The image in the focal plane D is the resulting coronagraphic DH image.} \label{fig:optical_layout}
\end{figure}

We use the PASTIS\cite{laginja2021analytical, laginja2022wavefront,leboulleux2018pair} approach to evaluate the maximum allowable segment-level surface deformations, and relate them to temperature requirements for individual segments. We use this approach to express the image plane DH intensity as a quadratic function of the WFE on the PM. This assumes that a nominal DH has already been created using static coronagraphic masks and a dynamic WFS\&C algorithm. In this scenario, PASTIS studies the impact of small phase aberrations on the spatial average contrast inside a DH. Note, that the contrast here is defined as the ratio of the coronagraphic image intensity at a point in the PSF and the peak intensity of the unocculted star. Whereas, the average DH contrast refers to a spatially average contrast across all the spatial resolution elements in the DH for an exposure. We use the finite element thermal models (shown in Fig.~\ref{fig:thermal_basis}) as the basis set of our analysis, and generalize the PASTIS analysis to accommodate multiple types of such segment-level aberrations. In particular, we build a matrix version for forward optical propagation using the thermal modes, with this matrix representation, we relate the average DH contrast to small wavefront errors in the pupil. The diffractive optical model we consider here consists of a segmented PM, a symmetric APLC binary pupil-plane mask, a circular focal plane mask and a circular Lyot stop. A schematic representation of the optical propagation is shown in Fig.~\ref{fig:optical_layout}. Below we summarize key mathematical formulas established in Leboulleux et al. (2018) and Laginja et al. (2021), and adapt these equations to incorporate the thermo-mechanical FEMs (in Fig.~\ref{fig:thermal_basis}) of the segment. 

For a small phase aberration $\phi(\vb{r})$ in the pupil aperture, the electric field $E(\vb{r})$ in the pupil plane is expressed as:
\begin{equation}
\begin{split}
  E(\vb{r}) & \simeq P(\vb{r}) + i\phi(\vb{r}),
\end{split}
\end{equation}
where $\vb{r}$ is the pupil-plane coordinate, $P(\vb{r})$ is the pupil function and $i$ the imaginary unit. Here, we have not considered any amplitude aberrations in the pupil and the small phase aberration is defined only over the pupil i.e. $\phi(\vb{r})$ is zero when $P(\vb{r})$ is zero.
Let $\mathcal{C}$ be a linear coronagraph operator, in which case the spatial intensity distribution $I_{f}$ in the focal plane is computed as:
\begin{equation}
\begin{split}
  {I_{f}}(\vb{u})
  &= |\mathcal{C}\{P(\vb{r})\} + i\mathcal{C}\{\phi(\vb{r})\} |^2  \\
  &= |\mathcal{C}\{P(\vb{r})\}|^2 + 2\mathfrak{Re}({\mathcal{C}\{P(\vb{r})\}\mathcal{C}\{\phi(\vb{r})\}^{*}}) +|\mathcal{C}\{\phi(\vb{r})\}|^2,
\end{split}
\label{eqn:I-full-form}
\end{equation}
where $\vb{u}$ is the focal-plane coordinate, $\mathfrak{Re}$ designates the real part and $*$ the complex conjugate.
In this case, where we have a symmetric DH, the cross-term is zero (the derivation is provided in Eqs. 33-36 in the Appendix of Leboulleux et al. (2018)), and the equation is reduced to:
\begin{equation}
\begin{split}
{\langle I_{f} \rangle}_{DH} &= {\langle{|\mathcal{C}\{P(\vb{r})\}|^2}}\rangle_{DH} + {\langle|i\mathcal{C}\{\phi(\vb{r})\}|^2\rangle}_{DH} \\
 &= c_{0} + {\langle|i\mathcal{C}\{\phi(\vb{r})\}|^2\rangle}_{DH}.
\label{eqn:I-and-c0-averaged}
\end{split}
\end{equation}
where $c_{0}$ is the nominal, static spatial average DH contrast limited by the design of the physical coronagraphic masks, ${\langle|i\mathcal{C}\{\phi(\vb{r})\}|^2\rangle}_{DH}$ is limited by speckles formed due to small wavefront perturbations and ${\langle \cdots \rangle}_{DH}$ refers to a spatial average over the DH. The phase $\phi(\vb{r})$ over the entire pupil can be expressed as a sum of the localized segment-level aberrations. Since we are considering only thermal effects, each of the segment-level aberrations is represented using the thermal basis \cite{hutterer2018advanced} as:
\begin{equation}
    \begin{split}
       \phi(\vb{r}) &= \sum_{(k,l)=(1,1)}^{n_{seg},5} a_{k,l}H_{l}(\vb{r}-\vb{r_{k}}),
       \label{eqn:phi-modes}
    \end{split}
\end{equation}
where $H_{l}(\vb{r}-\vb{r_{k}})$ represents the polynomial form of the surface deformation in Fig.~\ref{fig:thermal_basis}, $a_{k,l}$ represents the amplitude of the aberration for the $k_{th}$ segment for the $H_{l}$ mode of surface deformation and $\vb{r_{k}}$ is the coordinate of the center of the $k_{th}$ segment and following the algebra in Laginja et al. (2021), we find the spatial average DH contrast $c$ :
\begin{equation}\label{eqn:simple_matrix}
    \begin{split}
        c = c_{0} + \vb{a}^{T}M\vb{a},
    \end{split}
\end{equation}
where $\vb{a}$ is the pupil-plane aberration represented in the form of a 1d column vector, $M$ is defined as the PASTIS matrix, where each of its element is expressed as follows:
\begin{equation}\label{eqn:matrix_element}
\begin{split}
    m_{k,l,k',l'} &= {\bigg \langle  \mathcal{C}\bigg \{H_{l}(\vb{r}-\vb{r_{k}})
     \bigg \}  \mathcal{C}\bigg \{ H_{l'}(\vb{r}-\vb{r_{k'}})
     \bigg \}^{*} \bigg \rangle}_{DH}. \\
\end{split}
\end{equation}
For simplicity, we reduce the four-dimensional index of $m_{k,l,k',l'}$ to a two-dimensional index as: 
\begin{equation}\label{eqn:matrix_element}
\begin{split}
    m_{i,j} 
     &= {\langle \mathcal{C}\{\vb{e}_{i} \} \mathcal{C}\{{\vb{e}_{j}} \}^{*}\rangle}_{DH,}
\end{split}
\end{equation}
where either of $\vb{e}_{i}$ or $\vb{e}_{j}$ denotes a pupil aberration on a segment due to one of the five thermal modes and $i, j \in [1, nseg \times L]$, where $nseg$ is the total number of segments in the PM, $L$ is total number of thermal modes. Here, $m_{i,j}$ is a measurable quantity as it represents the change in the average contrast inside the DH due to both $\vb{e}_{i}$ and $\vb{e}_{j}$ perturbations in the pupil plane. 
From Eq.~\ref{eqn:simple_matrix}, we calculate the statistical mean contrast $\langle c\rangle$ of the DH spatial-averaged contrast over a normal distribution of $\vb{a}$ as:
\begin{equation}\label{eqn:variance}
    \langle c\rangle = c_{0} + \text{tr}(MC_{a}),
\end{equation}
where we define $C_{a} \equiv \langle \vb{a}\vb{a}^T \rangle$ as a variance matrix of the order $nseg\times L$. $C_{a}$ contains information about the thermo-mechanical correlations between the segments. The matrices $M$ and $C_{a}$ together describe the segment-surface response of the coronagraphic diffractive model to temperature disturbances on the PMSA. The PASTIS approach is also valid for a non-symmetric DH where the linear term in Eq.~\ref{eqn:I-full-form} can be eliminated by solving for the aberration vector $\vb{a}$ that provides the minimum contrast $c$, and then re-writing the equation under an appropriate change of variable into a quadratic expression similar to Eq.~\ref{eqn:simple_matrix} (the derivation is provided in the Eqs. 4-12 in Laginja et al. (2021)). Since we are using the statistical mean of the spatially averaged DH contrast to set stability requirements, in this scenario even if the linear term is considered, its statistical mean is constant and can be absorbed into $c_{0}$ in the tolerancing equations. Following Eq. 11 in Laginja et al. (2021),
\begin{equation}\label{eqn:linear_term}
\begin{split}
     \langle c\rangle &= c_{0} + \langle V^{T} \vb{a}\rangle + \langle \vb{a}^{T}M\vb{a} \rangle \\
     & = c_{0} + V^{T}\mu + \langle \vb{a}^{T}M\vb{a} \rangle \\
     & = c'_{0} + \langle \vb{a}^{T}M\vb{a} \rangle, \\
\end{split}
\end{equation}
where $c'_{0}$ is the effective constant term, $\mu$ is statistical mean of the aberration. The statistical mean metric is widely used in high-contrast imaging testbeds to quantify the performance of coronagraph.\cite{seo2019testbed} However, in scientific observations, the variance of contrast is the main metric used to assess the coronagraph’s detection sensitivity. Instead of statistical mean contrast, the variance of contrast may also be used to set the stability requirements\cite{steiger2026incorporating}, and is beyond the scope of this study. 
Under independent segment assumptions, $C_{a}$ is a diagonal matrix, thus Eq.~\ref{eqn:variance} is reduced to:  
\begin{equation}\label{eqn:mu_k}
    \langle c\rangle = c_{0} + \sum_{k = 1}^{nseg*L} m_{kk}\mu^2_{k},
\end{equation}
where $m_{kk}$ is the diagonal element of the PASTIS matrix $M$ and $\mu_{k}$ is the standard deviation of the wavefront aberration on $k_{th}$ segment. Eq. \ref{eqn:mu_k} is based on the assumption that, for each FEM mode considered separately, errors across different segments are mutually uncorrelated, resulting in diagonal covariance matrices across segments, and then stacked into a single covariance matrix. Coupling between the FEM modes is not included in this analysis, which may give rise to a non-diagonal covariance matrix. The tolerancing can be done by eigen decomposition on the non-diagonal covariance matrix, which gives a set of independent modes that describes the system’s thermo-mechanical response. The mathematical derivation is provided in Section 4.3 of Laginja et al. (2021). The PASTIS analysis can be used to allocate uniform contrast across all the FEM modes, which further simplifies Eq.~\ref{eqn:mu_k} to:
\begin{equation}
    \langle c\rangle = c_{0} + nseg*L\times m_{kk}\mu^2_{k}
\end{equation}
and 
\begin{equation}\label{eqn:mu_k2}
    \mu_{k} = \sqrt{\frac{\langle c\rangle - c_{0}}{nseg*L\times m_{kk}}}.
\end{equation}
Eq.~\ref{eqn:mu_k2} allows us to calculate the maximum allowable surface deformation per thermal mode, per segment, required to maintain a desired statistical mean-contrast stability inside the DH over multiple realizations of wavefront error maps built by the finite element thermal models. 

\subsection{Results of static tolerancing}
\begin{figure}[!ht]
\centering
\includegraphics[width=0.8\linewidth]{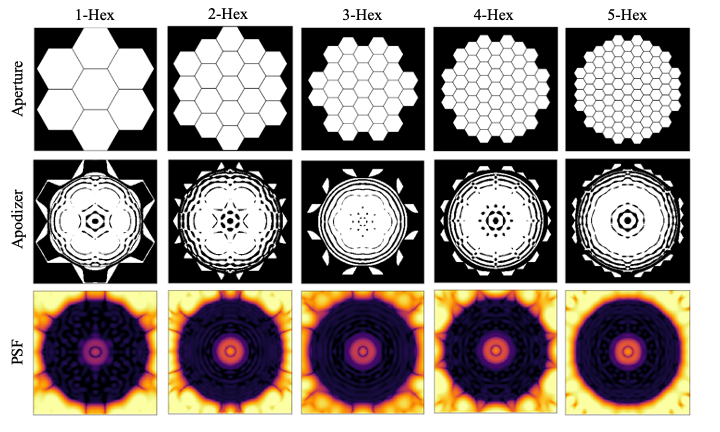}
 \caption{\textit{Top:} Hexagonal segmented primary mirror designs with no central obscuration with different number of rings and varying segment sizes (flat-to-flat (m) 1-Hex: 2.64, 2-Hex: 1.45, 3-Hex: 1.19, 4-Hex: 0.85, 5-Hex: 0.66), all designs have an inscribed diameter of $\sim$6 m. \textit{Middle:} Binary apodizer masks optimized for their respective aperture design. \textit{Bottom}: Respective coronagraphic PSFs with an annular dark hole ranging from 3.4$\lambda/D$ - $12\lambda/D$, and a spatial average contrast floor of $\sim$$4\times 10^{-11}$ inside the dark hole.} \label{fig:all_hex3}
\end{figure}

For a comparative study of WFE tolerances with different segment sizes and numbers, we use the multiple segmented PM architectures with their optimized coronagraph computed by the SCDA study, which was organized by NASA's Exoplanet Exploration Program (ExEP) to provide coronagraph designs for possible future obscured and unobscured segmented apertures. The designs we use for our analysis is the case of a $\sim$6~m aperture off-axis segmented aperture\cite{nickson2022aplc}. The PM architectures consist of hexagonal mirror segments arranged in N = 1, 2, 3, 4 and 5 concentric rings around a central segment with no obscuration. All of these designs have an inscribed diameter of $\sim$6 m and 6 mm segment gaps, with varying segment sizes (flat-to-flat (m) 1-Hex: 2.64, 2-Hex: 1.45, 3-Hex: 1.19, 4-Hex: 0.85 and 5-Hex: 0.66). Respective apodizer solutions for these designs were obtained using the publicly available APLC-Optimization toolkit\cite{aplc_optimization}, which optimizes the apodizer design for a given focal-plane mask and Lyot stop geometry. The SCDA apodizer designs were optimized for a spatial average DH contrast floor below $10^{-10}$ and for a $10\%$ spectral bandwidth. A schematic representation of the 5 types of PM geometries with their optimized APLC coronagraphs is shown in Fig.~\ref{fig:all_hex3}. The apodizer solutions provide similar throughput and spatial average contrast floor inside the dark region. More details on the geometry specifications, apodizer optimization method, core throughputs, robustness of these designs to optical misalignment are described in Nickson et al. (2022)\cite{nickson2022aplc}.

\begin{figure}[!ht]
\centering
\includegraphics[width=1.06\textwidth, height=0.78\textheight, center]{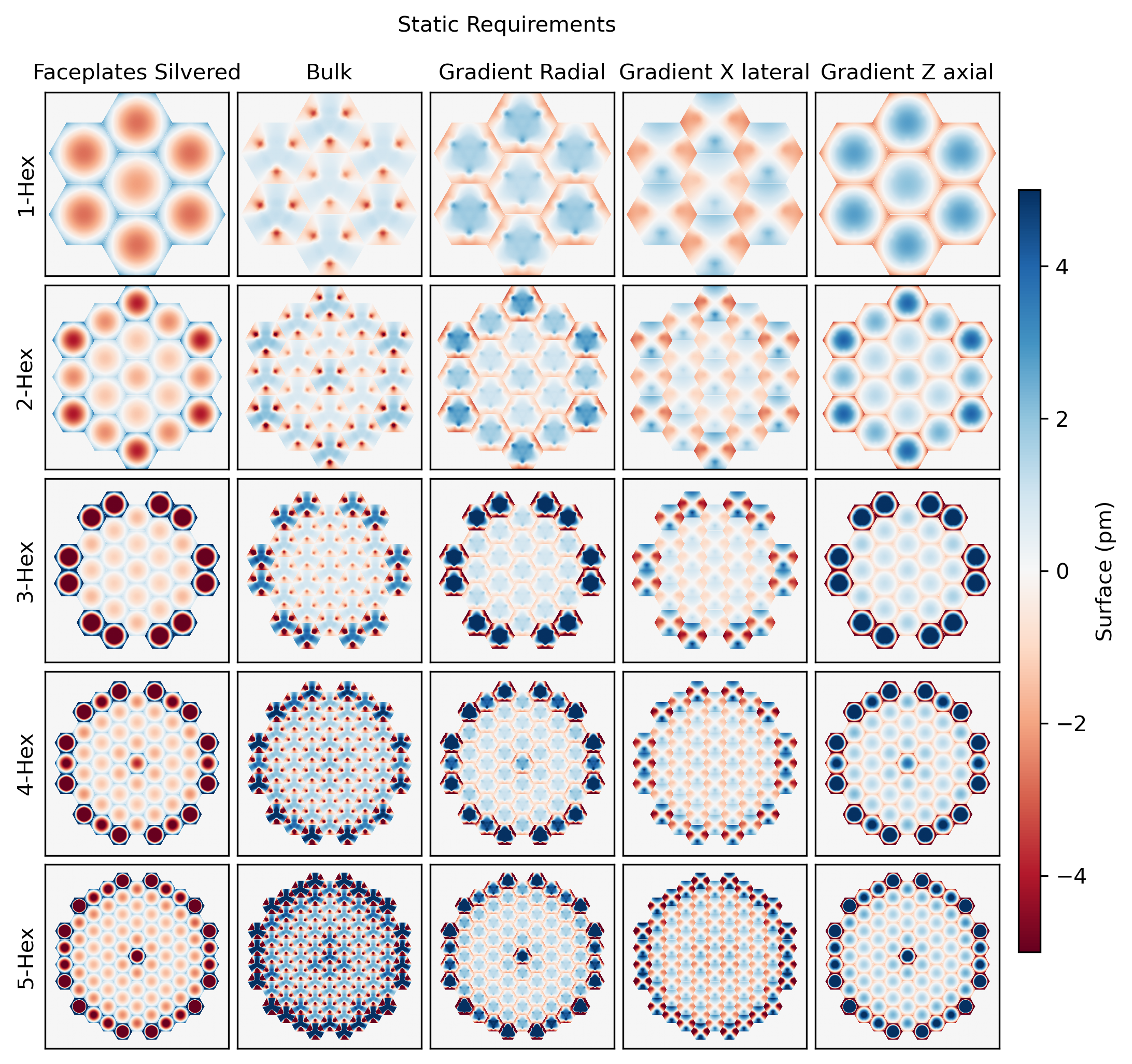}
\caption{Segment-level allowable static surface deformations for the five SCDA designs, so that the raw contrast does not differ from the design contrast by more than $10^{-11}$, at a wavelength of 500 nm. With increase in segment numbers (or decrease in segment size), we observe that the outer segments can withstand large surface deformation compared to the inner segments.} \label{fig:all_hex_harris_pm}
\end{figure}

\begin{figure}[!ht]
\centering
\includegraphics[width=1.06\textwidth, height=0.78\textheight, center]{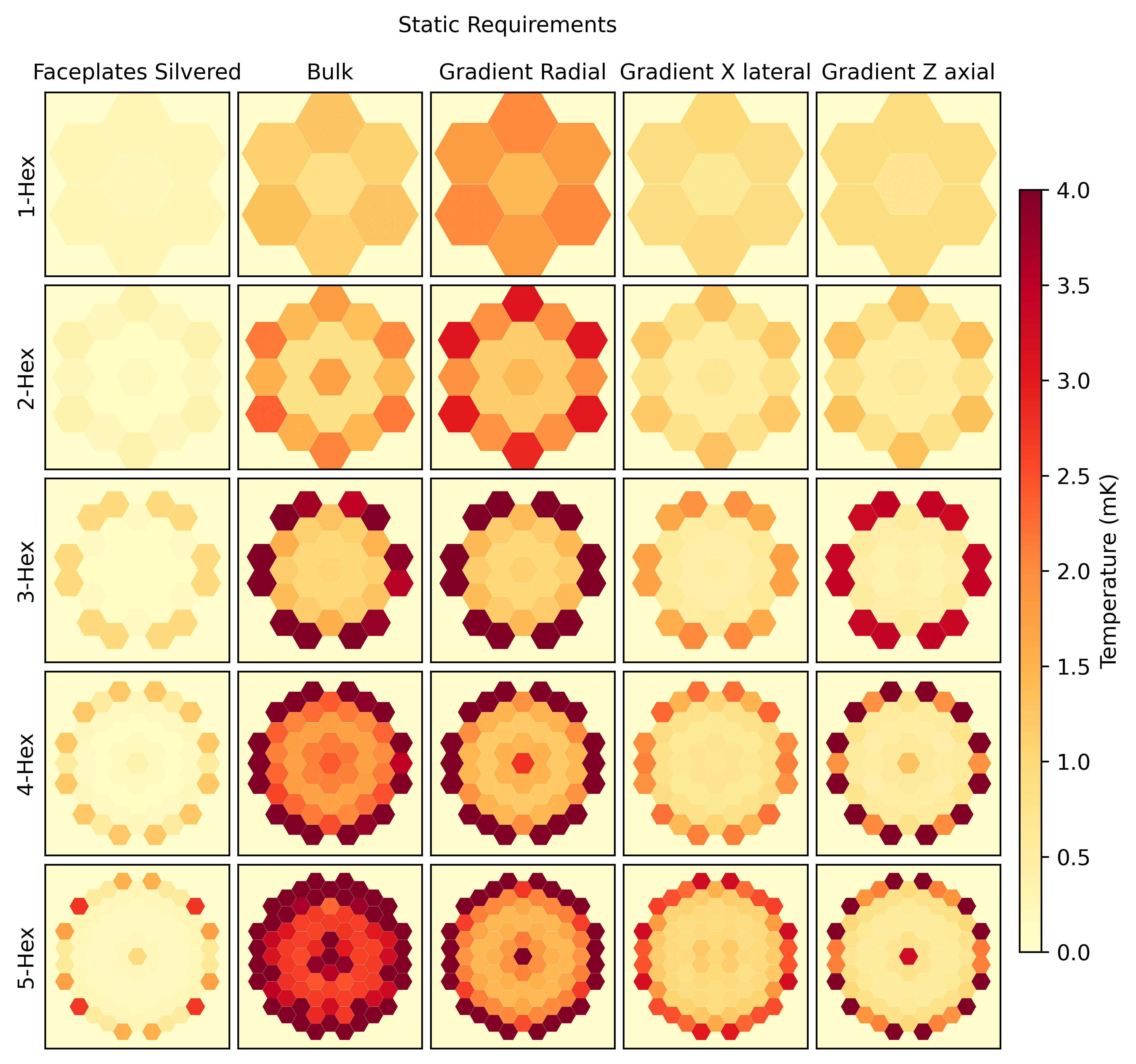}
\caption{Segment-level allowable static temperature requirements in mK for the five SCDA designs, so that the raw contrast does not differ from the design contrast by more than $10^{-11}$, at a wavelength of 500 nm. To calculate the temperature requirements, we first obtained the amplitudes of surface tolerances represented by the FEMs in Fig.~\ref{fig:all_hex_harris_pm}, and then converted these amplitudes to mK scale using the temperature-gradient relation in Fig.~\ref{fig:thermal_basis}.} \label{fig:all_hex_static_mK}
\end{figure}

To calculate the static requirements, we first calculate the contrast-based PASTIS matrix, and then use the diagonal elements of this matrix to determine the segment-level tolerances. To build the PASTIS matrix, we poke a pair of segments with one of the five thermal modes, and propagate the resulting wavefront arising from the surface deformation through a diffractive optical model of the SCDA mirror architecture with their coronagraph mask. We then obtain the spatial average DH contrast due to the aberrated segment pairs and subtract it from a reference contrast floor (i.e. $c_{0}$, corresponding to the case of no external aberration). We use these change-in-contrast measurements to fill all the elements of the PASTIS matrix in the order described in Laginja et al. (2021)\cite{laginja2021analytical}. The segment-level static surface tolerances, $\mu_{k}$s for the five types of telescope geometries, to maintain a spatial average DH contrast stability of $c = 10^{-11}$ for all five thermal modes is computed using Eq.~\ref{eqn:mu_k2}. We map these results onto individual segment-surfaces as shown in Fig.~\ref{fig:all_hex_harris_pm}, illustrating non-uniform requirements across the PMs. Note, these assume equal weighting of contrast aberration, and could be adjusted in the design allocation phase. The magnitude of $\mu_{k}$ on each segment represents the standard deviation of a zero-mean normal distribution, or alternatively maximum permissible surface error on the segment. In general, we notice the outer rings of all designs have higher tolerances than the inner rings. The requirement for each segment is dependent on the geometrical design of the APLC masks, and is related to the throughput received after the apodized pupil plane. The segments which are significantly blocked by the mask have less impact on the coronagraphic PSF and thus have loose stability requirements. Also, for a given telescope geometry, the requirements are different for different types of FEMs. A detailed quantitative chart showing individual segment requirements for the five SCDA designs, for a single thermal load is shown in Table~\ref{tbl:num0_static_table}. We find that the tolerances also depend on the number of segments in the pupil. Wavefront aberrations arising from mirrors with fewer segments typically fall in the regime of low spatial frequencies and are filtered by the coronagraph. This leads to higher tolerances for pupil with fewer segments. This is evident as we find that the 1-Hex design has a higher RMS tolerance compared to the RMS tolerance for the inner segments of the other designs. Inversely, with more segments, the wavefront aberrations represented by the FEMs fall in the high spatial frequencies regime and are less filtered by the coronagraph. The inner rings of the 3-Hex design have lower tolerances than those of the 2-Hex design. Overall, we find that the number of segments in the pupil creates competing effects on the extent of pupil apodization. The picometer-tolerance maps are then scaled to mK-tolerance maps using the FEMs (see Fig.~\ref{fig:thermal_basis}) which relates surface deformation for a calibrated thermal stress. We obtain individual segment-level temperature requirements by dividing the static surface-tolerance maps with the mK-unit surface maps of FEMs. In Fig.~\ref{fig:all_hex_static_mK}, we plot the thermal standard deviation of a zero-mean normal distribution for each segment for all five telescope geometries. Consistent with the static picometer-level tolerance analyses, we obtain the non-uniform temperature requirements across all segments for the five SCDA designs, and they depend on the nature of the thermal loads applied to the segment as well as on the optical design architecture. We further validate the picometer-tolerances with end-to-end Monte-Carlo (MC) simulations, where a segment-level aberration map is drawn from a zero-mean normal distribution with standard deviation as $\mu_{k}$ and is propagated through the optical model. We then register the contrast stability due to this aberration map. Table~\ref{tbl:static_table} shows the mean spatial contrast stability allocations inside the DH over multiple realizations of wavefront error drawn from the distribution. The RMS contrast stability across all the five geometries shown in the table is in close agreement with the desired contrast stability of $10^{-11}$. Note, the PASTIS approach is limited to defining only the static tolerances for the segments and it does not specify how these tolerances are to be maintained during a long science observing scenario. The values presented in Table~\ref{tbl:num0_static_table} and \ref{tbl:static_table} are applicable to the case where the observatory's environmental condition does not change over time, i.e. the pupil plane small aberration is static and there is no dynamic evolution of speckles inside the DH. In the presence of dynamic speckles, the average mean intensity changes will also change with time, thus changing the tolerance requirements computed from Eq.~\ref{eqn:mu_k2}. We address this issue in \S~\ref{sec:temporal}. 

\begin{table}[!ht]
\centering
  \includegraphics[height = 0.35\textheight, width=1.01\linewidth, center]{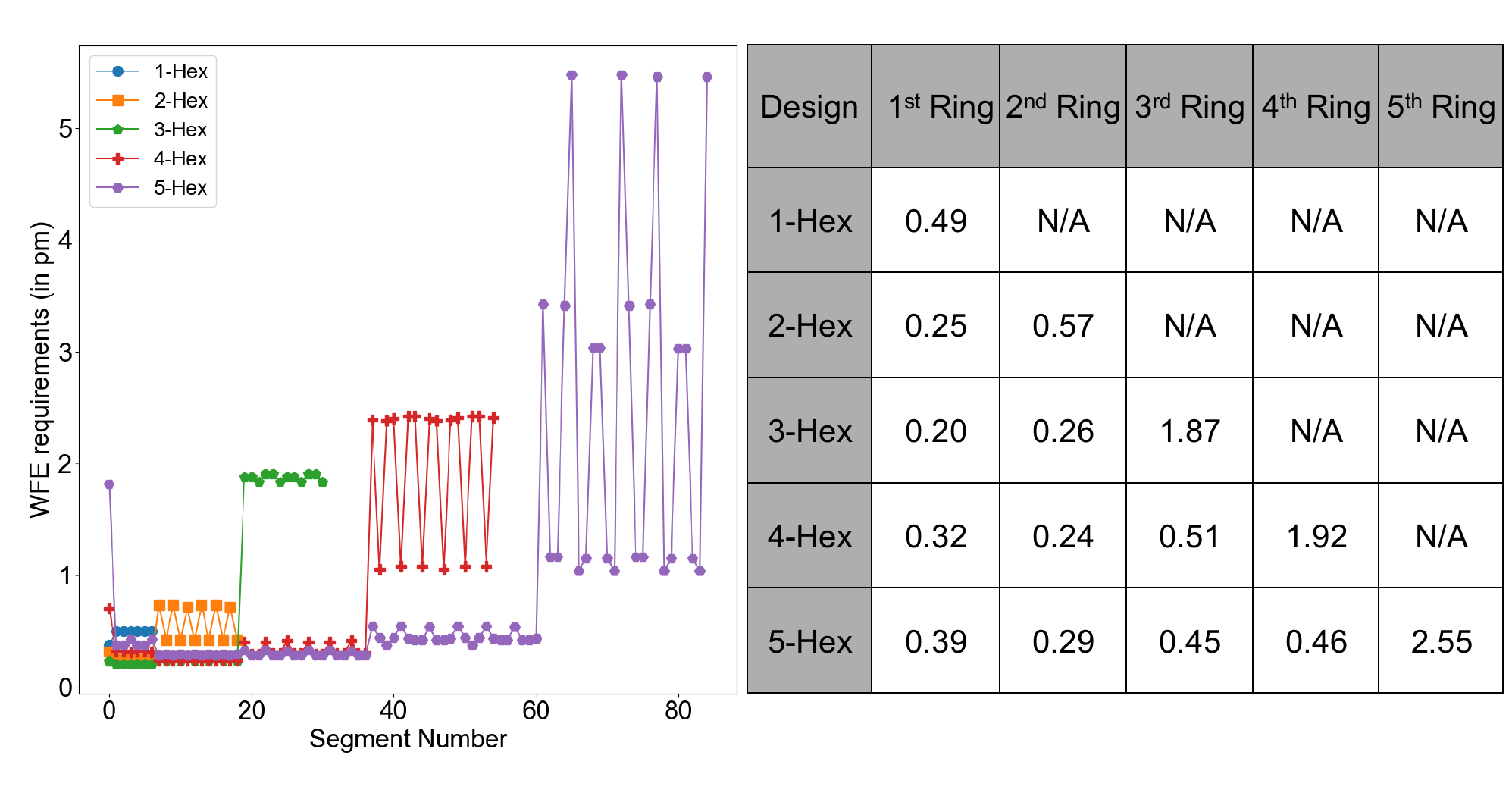}
  \caption{\textit{Left:} Static wavefront stability requirements for individual segments, so that the raw contrast does not differ from the design contrast by more than $10^{-11}$, for a finite element thermal model (Faceplates Silvered). \textit{Right:} Corresponding RMS static wavefront stability allocation (in units of picometer) for each of the rings of the PM across all of the SCDA designs.}\label{tbl:num0_static_table}
\end{table}

\begin{table}[!ht]
\centering
\caption{Top-level contrast error budget for the SCDA designs in the presence of static wavefront error due to the segment-level thermal gradients}
\begin{subfigure}{0.47\textwidth}
    \centering
    \includegraphics[page=1, height = 0.35\textheight, width=\linewidth]{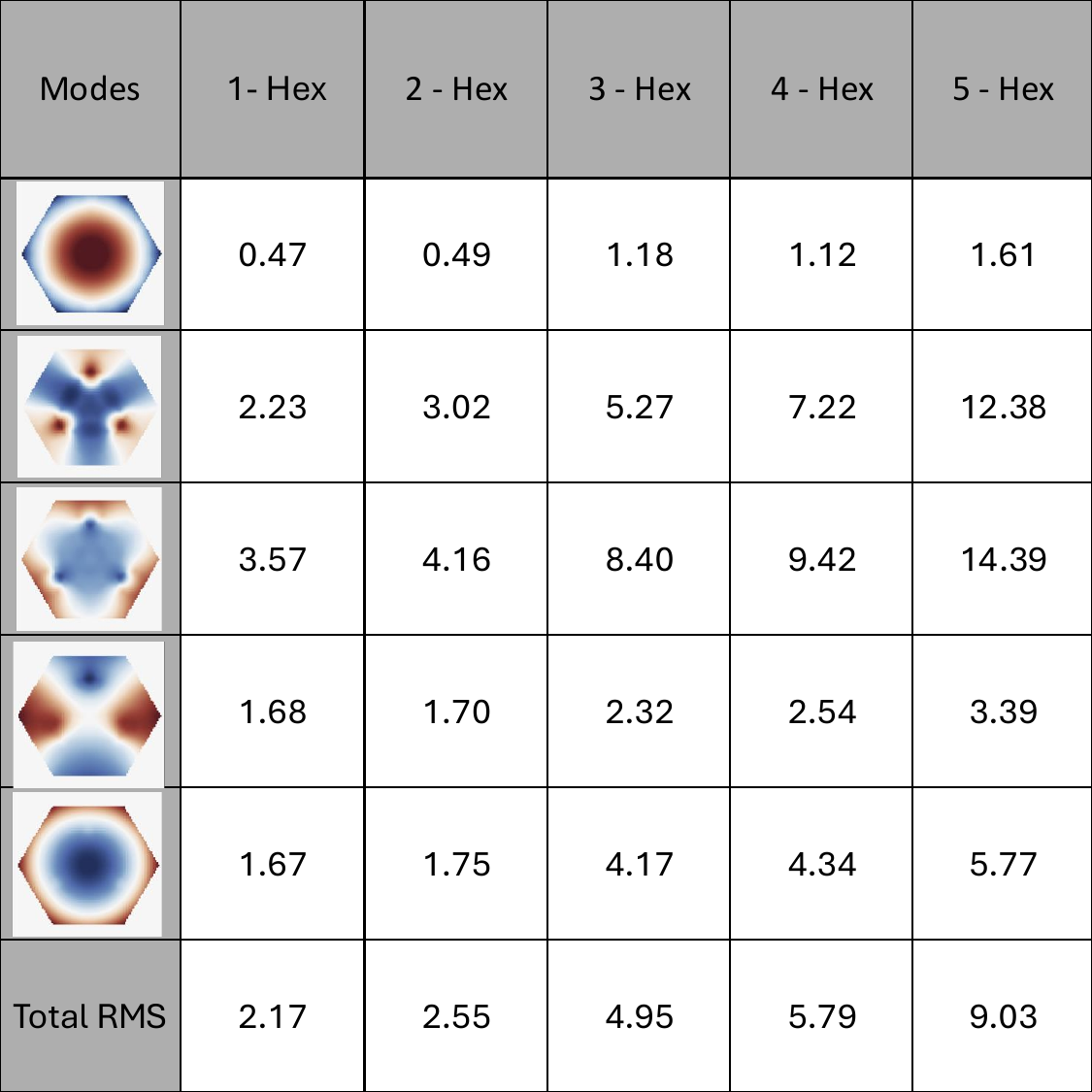}
    \caption{Tolerances}
    \end{subfigure}
    \hspace{0.005\textwidth} 
    \begin{subfigure}{0.47\textwidth}
    \centering
    \includegraphics[page=2, height = 0.35\textheight, width=\linewidth]{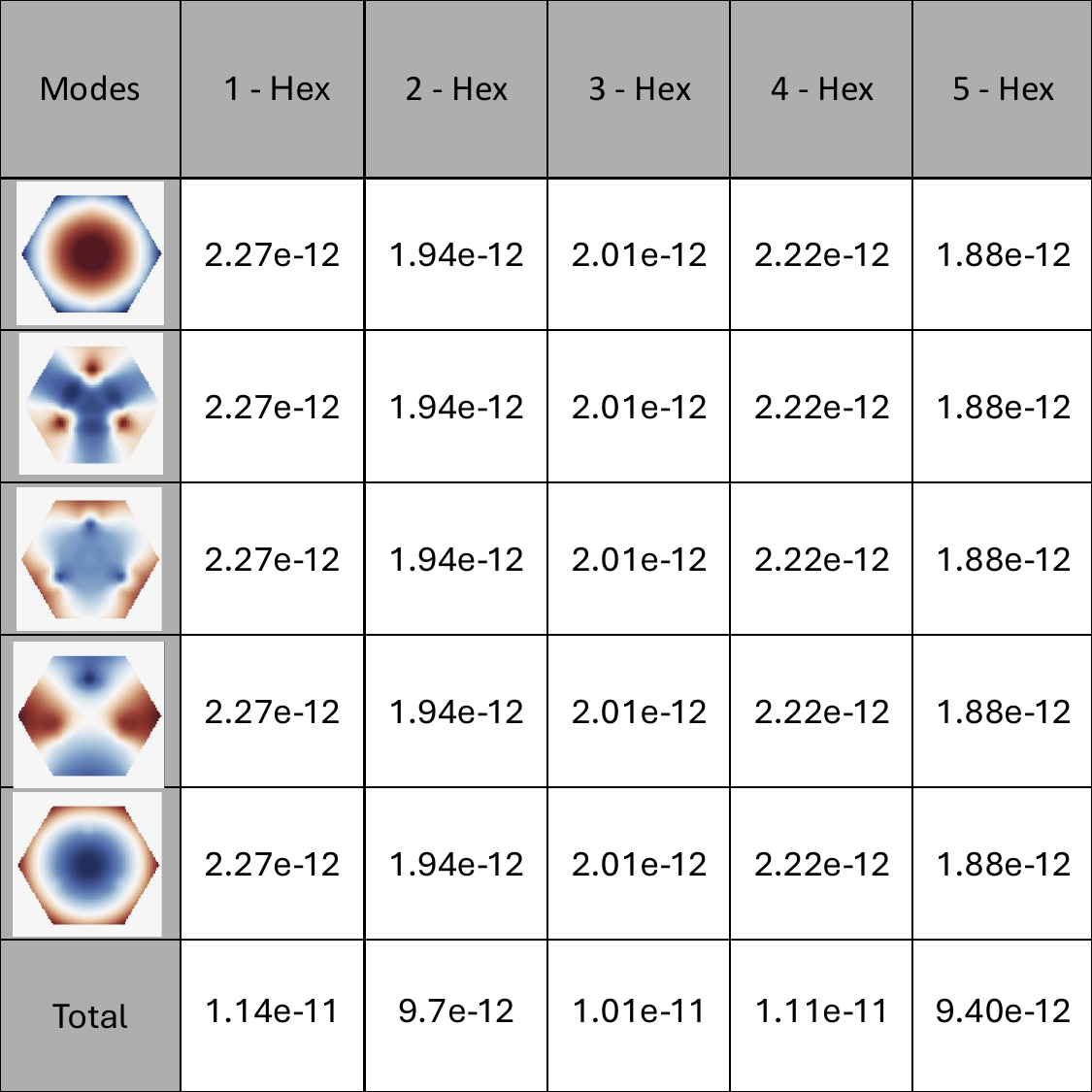}
    \caption{Contrast Allocation}
    \end{subfigure}
  \caption*{\textit{Left:} RMS static wavefront error allocation (in units of picometer) across the PM of the SCDA designs, so that the raw contrast does not differ from the design contrast by more than $10^{-11}$. \textit{Right:} Respective raw static DH contrast deviations allocation for each of the thermal modes, obtained via end-to-end MC simulations, where a segment-level aberration map is drawn from a normal distribution with standard deviation $\mu_{k}$ and zero-mean, and then propagated through the optical model.} \label{tbl:static_table}
\end{table}
\clearpage

\section{Allowed temporal drifts for maintaining a difference of $\text{10}^{\text{--11}}$ between raw and design contrasts}\label{sec:temporal}
\subsection{Method}
In this section, we discuss the methodology for implementing an information-theoretical approach to estimate allowable dynamic WFEs in a linear regime of small wavefront perturbation, assuming a WFS\&C algorithm is used to maintain the design contrast.
To do so, we apply the discrete-time batch estimation algorithm established in Pogorelyuk et al. (2021)\cite{pogorelyuk2021information} to relate the open loop wavefront drifts to the closed loop residuals. We choose our spatial modes as  scaled up or down versions of the static tolerances maps obtained in \S~\ref{sec:pastis}, and iterate these modes in the batch-estimation algorithm. This method yields a lower bound on the allowable dynamic surface deformation, subsequently on the thermal deviations. We estimate a lower bound on the wavefront variances based on the Cramer-Rao inequality\cite{cramer1946contribution, rao1992information}, i.e. we compute the Fisher Information on the wavefront variances, then relate this information to DH contrast using integrated diffractive models of the telescope and coronagraph architectures, a low read-noise science camera and a non-common-path Zernike wavefront sensor. We first begin with summarizing the key mathematical formulation outlined in Pogorelyuk et al. (2021)\cite{pogorelyuk2021information} in the context of our study, and then use this method to estimate lower bounds on close-loop drifts. 

Let $\epsilon_{k}$ be the open-loop coefficients for localized spatial WFE modes for a $k_{th}$ exposure of a WFS\&C sequence. These WFE modes form a basis set that can be used to express an arbitrary pupil-plane wavefront error and we use the static tolerances derived in \S~\ref{sec:pastis}  as a modal basis in this section. We assume $\epsilon$ to follow a simple Brownian motion, thus increments of $\epsilon$ are normally distributed with zero mean and drift-variance denoted as Q,  
\begin{equation}\label{eqn:open-loop-estimates}
    \epsilon_{k+1} - \epsilon_{k} \sim \mathcal{N}(\textbf{0}, Q), \; Q>0,
\end{equation}
Let $\hat{\epsilon}$ be the unbiased or true estimate of the coefficient of the WFE modes by estimators (such as a wavefront sensor), and its error is also assumed to be normally distributed with a variance $P_{k}$, 
\begin{equation}\label{eqn:unbiased-estimates}
    \hat{\epsilon_{k}} - \epsilon_{k} \sim \mathcal{N}(\textbf{0},P_{k}),       \; P_{k} > 0.
\end{equation}
Let $\epsilon^{CL}$ be closed-loop estimate of the coefficients of the WFE modes, and we relate it to the open-loop estimates as:
\begin{equation}\label{eqn:close-loop-relation}
\begin{split}
\epsilon^{CL}_{k+1} = \epsilon_{k+1}-\hat{\epsilon_{k}}.
\end{split}
\end{equation}
Substituting the expression for $\epsilon_{k+1}$ and $\hat{\epsilon_{k}}$ in Eq.\ref{eqn:close-loop-relation}, we infer that $\epsilon^{CL}$ is also normally distributed, and is expressed as:
\begin{equation}\label{eqn:drift_model}
    \begin{split}
        \epsilon^{CL}_{k+1} \sim \mathcal{N}(\textbf{0}, P_{k} + Q).
    \end{split}
\end{equation}
We assume that the `steady-state' electric field $\textbf{E}$ at the image plane, once the nominal DH has been created, is a linear function of $\epsilon^{CL}$ and can be expressed as:
\begin{equation}
    \textbf{E} = G\epsilon^{CL} + \textbf{E}_0 ,
\end{equation}
where we define $G$ to be a sensitivity matrix and $\textbf{E}_{0}$ to be a static, uncontrolled reference electric field without any aberrations. Here $G$ holds information about the electric fields to the known wavefront aberrations. The intensity $I$ at the image plane is
\begin{equation}\label{eqn:intensity}
I = \dot{N_{S}}||G\epsilon^{CL} + \textbf{E}_{0}||^2,
\end{equation}
$\dot{N_{S}}$ is defined as the photon throughput at the image plane, and the electric field $\textbf{E}$ is normalized to one photon in the pupil. 
We use the analytical expression derived in Pogorelyuk et al. (2021) to relate $\textbf{E}$ and Fisher information $\mathcal{I}$, and is expressed as follows:
\begin{equation}\label{eqn:fisher}
    \mathcal{I} = \frac{4\dot{N_{S}}t_{s}}{||G\epsilon^{CL} + \textbf{E}_{0}||^2+\dot{N}^{-1}_{S}(D^{n})}G^{T}(G\epsilon^{CL} + \textbf{E}_{0})(G\epsilon^{CL} + \textbf{E}_{0})^{T}G,
\end{equation}
where $D^{n}$ represents total noise from incoherent external (such as zodiacal dust) and internal sources (clock-induced charge, dark current), $t_{s}$ represents detector integration time. Eq.\ref{eqn:fisher} represents an analytical method to compute Fisher information $\mathcal{I}$ from close-loop coefficients $\epsilon^{CL}$ of WFE modes. Using Cram\'er-Rao inequality\cite{cramer1946contribution}, we further relate the distribution of $\epsilon^{CL}$ with $\mathcal{I}$ as:
\begin{equation}\label{eqn:cramer-rao}
\begin{split}
    var(\hat{\epsilon^{CL}}) \geq \frac{1}{\mathcal{I}(\epsilon^{CL})},
\end{split}
\end{equation}
and based on the this formulation, a lower bound on the close-loop variance represented by $P_k$ (see Eq. \ref{eqn:unbiased-estimates}) can be estimated as:
\begin{equation}\label{eqn:cramer-rao}
    P_{k} \geq \frac{1}{{\mathcal{I}_{k}}}.
\end{equation}
By using the probabilistic WFE drifts model (Eq. \ref{eqn:drift_model}) and the Fisher information measurement model (Eq. \ref{eqn:fisher}), we can compute $P_{k}$ for an exposure. In the steady-state regime, $P$ does not change much for the nearest exposure (i.e. $P_{k+1}\approx P_{k}\approx P$), and therefore the Fisher information can be averaged to a constant across all exposures. Finding a steady state solution for $P$ requires computing expectation of $\mathcal{I}$ w.r.t $\epsilon^{CL}$ which can be challenging to solve. Instead, we use the iterative batch-estimation algorithm based on random sampling described in Pogorelyuk et al. (2021). We first compute sensitivity matrices, $G$s at both the science and the wavefront sensor planes. These matrices are constructed just as the PASTIS matrix described in \S~\ref{sec:pastis}.
Scaled up or down segment-level static requirement values, $\mu_k^{2}$s, computed from the PASTIS analysis in \S~\ref{sec:pastis} serve as the open-loop wavefront error drifts variance, $Q$. Photon flux from a star, based on its apparent magnitude, spectral type, and instrument band pass is computed using the exoscene package\cite{exoscene_citation}, which includes tools for computing spectral irradiance using the Bruzual-Persson-Gunn-Stryker (BPGS) Spectral Atlas. This flux is further throughput-corrected using the aperture sizes and apodizers' transmittance. Using this information, we compute $\mathcal{I}$s and the respective close-loop contrasts over different wavefront sensing exposure times using the batch estimation algorithm. The final covariance estimate is obtained by averaging all $P$s over the iterations, and we also compute the corresponding converged mean contrasts using Eq.\ref{eqn:intensity}. Since our main goal is to set requirements on $Q$, which informs us how much the observatory can drift in an open-loop scenario, we iterate the batch estimation algorithm over a range of $Q$ values, and obtain the corresponding close-loop contrast-stabilites as outputs. We then ultimately choose and quote the $Q$ that yields a close-loop contrast-stability closet to our target contrast stability, $10^{-11}$.

\subsection{Results of dynamic tolerancing}
\begin{figure}
    \centering
     \begin{subfigure}{0.49\textwidth}
         \centering
         \includegraphics[page=1, height = 0.28\textheight, width=\linewidth]{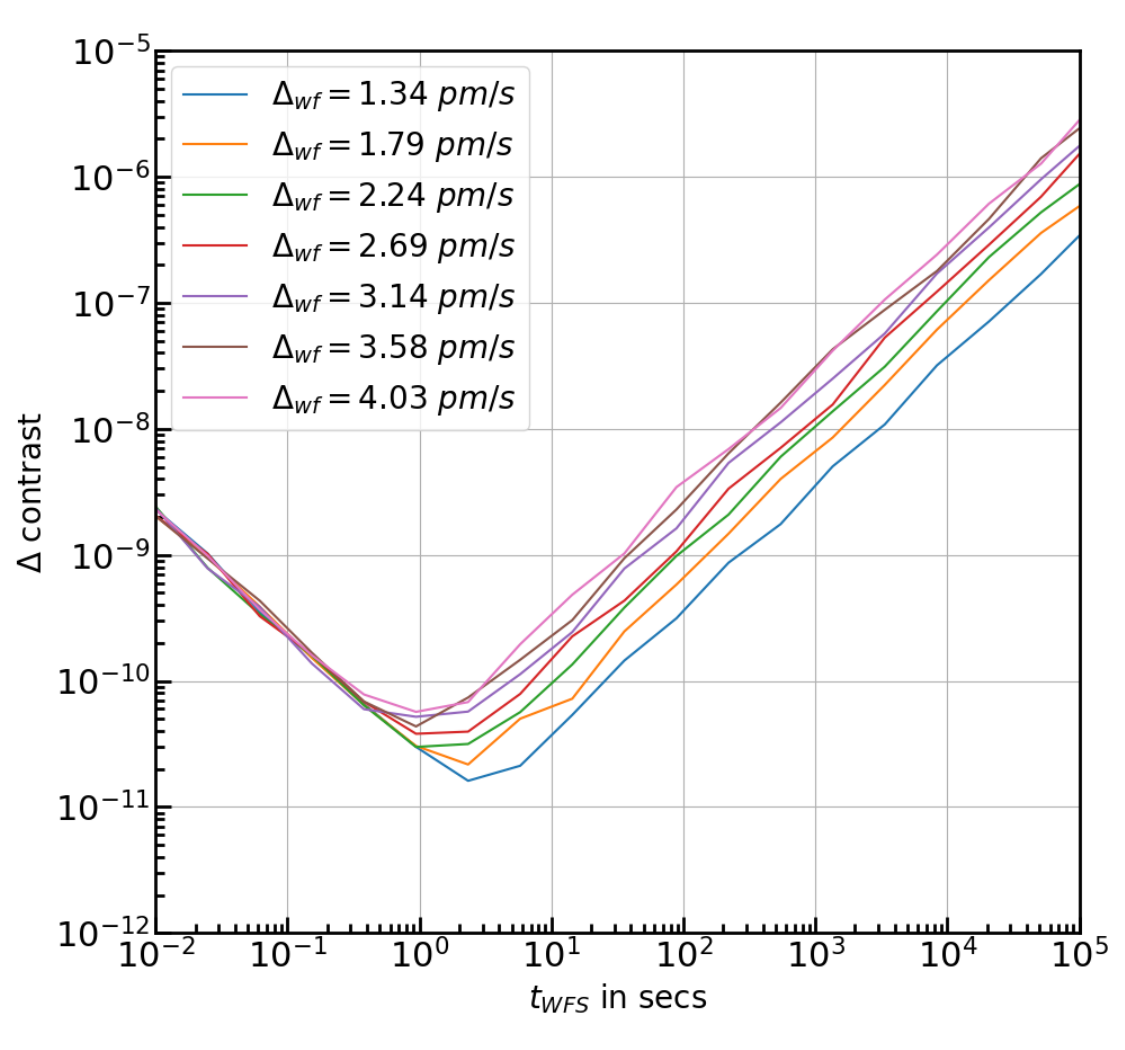}
         \caption{For 1-Hex design, comprising of 7 segments}
         \label{fig:1-Hex temporal}
     \end{subfigure}
     \begin{subfigure}{0.49\textwidth}
         \centering
         \includegraphics[page=2, height = 0.28\textheight, width=\linewidth]{contrast_wf_all_hex_v2.pdf}
         \caption{For 2-Hex design, comprising of 19 segments}
         \label{fig:2-Hex temporal}
     \end{subfigure}
     
     \vspace{0.1cm}
     
     \centering
     \begin{subfigure}{0.49\textwidth}
         \centering
         \includegraphics[page=3, height = 0.28\textheight, width=\linewidth]{contrast_wf_all_hex_v2.pdf}
         \caption{For 3-Hex design, comprising of 31 segments}
         \label{fig:3-Hex temporal}
     \end{subfigure}
     \begin{subfigure}{0.49\textwidth}
         \centering
         \includegraphics[page=4, height = 0.28\textheight, width=\linewidth]{contrast_wf_all_hex_v2.pdf}
         \caption{For 4-Hex design, comprising of 55 segments}
         \label{fig:4-Hex temporal}
     \end{subfigure}

     \centering
     \begin{subfigure}{0.49\textwidth}
         \centering
         \includegraphics[page=5, height = 0.28\textheight, width=\linewidth]{contrast_wf_all_hex_v2.pdf}
         \caption{For 5-Hex design, comprising of 85 segments}
         \label{fig:5-Hex temporal}
     \end{subfigure}   
        \caption{Close-loop contrast stability at different wavefront-sensor integration times for the five SCDA designs along with their optimal dynamic tolerances.}
        \label{fig:SCDA temporal}      
\end{figure}

\begin{figure}[hbtp]
\centering
\includegraphics[width=1.06\textwidth, height=0.78\textheight, center]{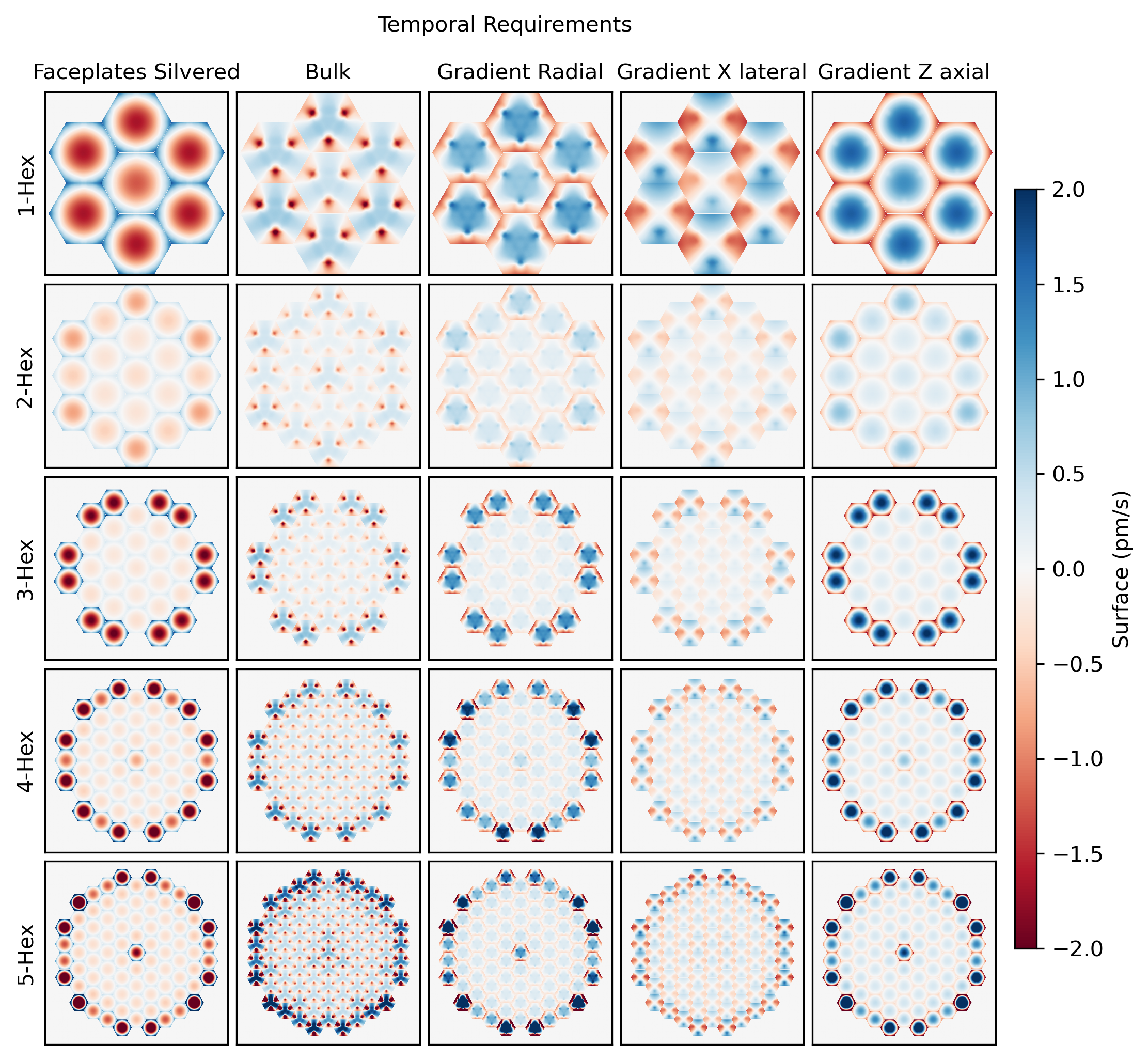}
 \caption{Segment-level maximum allowable dynamic surface deformation for the five SCDA designs to maintain a DH contrast stability of $10^{-11}$. The optimal wavefront error drift-scales obtained using the batch estimation algorithm are used to scale the static tolerance maps into dynamic tolerance maps.} \label{fig:all_hex_pm_m}
\end{figure}

\begin{figure}[hbtp]
\centering
\includegraphics[width=1.06\textwidth, height=0.78\textheight, center]{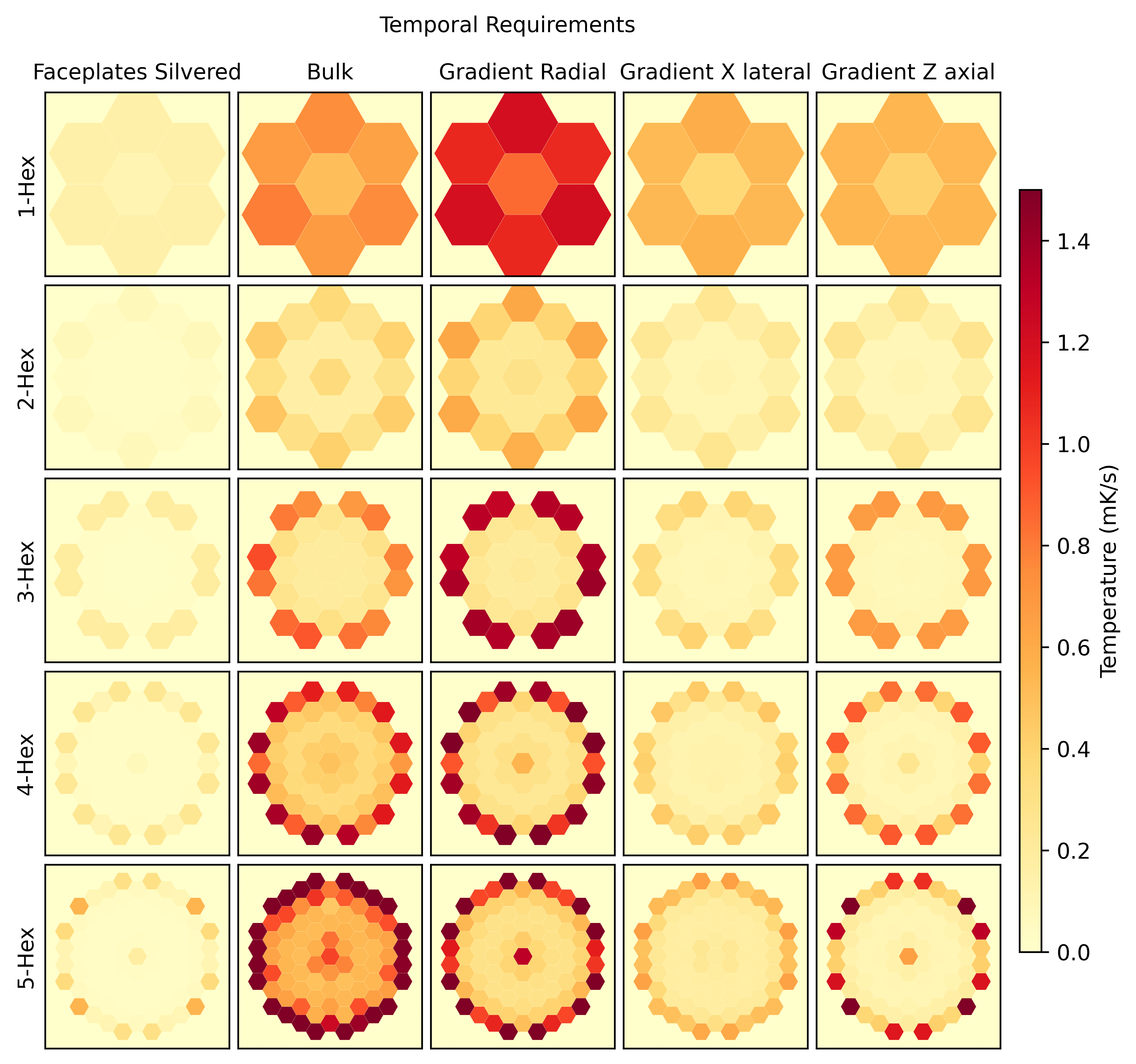}
 \caption{Segment-level maximum allowable dynamic thermal deviation for the five SCDA designs to maintain a DH contrast stability of $10^{-11}$. The optimal wavefront error drift-scales obtained using the batch estimation algorithm are first used to generate the dynamic surface tolerances, these surface tolerances are then translated to thermal tolerance maps using the FEMs shown in Fig.\ref{fig:thermal_basis}.} \label{fig:all_hex_mK_s}
\end{figure}

Using the batch estimation algorithm, for a $5^{th}$ magnitude star in the V band, assuming zero incoherent noise ($D^n$), we compute close-loop contrast stabilities at different WFS integration times, under a range of temporal external drift scenarios $Q$s. Fig.~\ref{fig:SCDA temporal} shows close-loop contrast stability at different wavefront-sensor integration times for the five SCDA designs across different temporal wavefront error drift scenarios. We observe a global minimum contrast in these curves. The minima in these curves represent the optimal wavefront sensing time scale for a temporal drift. On the left-hand side of the minimum, the contrast is limited by photon noise, while on the right-hand side, it is degraded due to the sensing exposure being too long to correct drifts on time. We select the scaling where the minimum corresponds to our desired contrast stability, in our case, $10^{-11}$. The corresponding drift is the largest possible open loop drift that still yields the required stability. The $\Delta_{wf}$s shown in Fig.~\ref{fig:SCDA temporal} are scaled versions of $Q$s. The list of $\Delta_{wf}$ scaling values is not necessarily the same for each subplot. We choose these scaling values to identify the $\Delta_{wf}$ scale that most closely achieves the desired contrast stability. We observe in this figure a largest allowable open loop drift of ${\sim}1-2$ pm/s across all SCDA designs required to maintain a close loop contrast of $10^{-11}$. In Fig.~\ref{fig:SCDA temporal}, we iterate over multiple scaled values of $Q$. Theoretically, all designs can be made to reach the exact contrast stability of $10^{-11}$ or lower than that, depending on the $Q$-scales chosen. We use a coarse grid here to find the approximate $Q$ for our desired stability requirements. In this grid, the 2-Hex design reaches $10^{-11}$ for a certain scaled version of $Q$; however, we do not use the corresponding scaled $Q$ ($\Delta_{wf}=0.52$ pm/s) for our comparative study. Instead we use the optimal values from the orange curves where all the designs reach approximately similar level of contrast stability (i.e. $\sim$1-2$\times 10^{-11}$). We use the optimal drifts and wavefront-sensing exposure times obtained from these iterations for subsequent contrast error budgeting. These $\Delta_{wf}$ drifts are the mean of `drifts' across all segments, and do not represent individual requirements for a segment. For a comparative study across all the segments, we then display the results on a segment to segment basis in Fig.~\ref{fig:all_hex_pm_m} and ~\ref{fig:all_hex_mK_s}. The temporal tolerances across the segments follow similar patterns to those we find in the static analysis. This is because we use the scaled versions of the PASTIS tolerances, which represent the tolerances across individual segments. The segments that are partially or completely blocked by the apodizer have higher temporal tolerances than the ones that are unblocked. Our analysis shows that the 5-Hex design, comprising 85 segments, can withstand higher mean $\Delta_{wf}$ drift (${\sim}2$ pm/s) compared to other SCDA architectures to achieve similar scientific goal. We also find that the 1-Hex design can achieve the close-loop contrast stability much faster than the other designs. This is because the collecting area of each segment is larger, thus enabling more information rich sensing exposures. This operational advantage of larger segment has to be traded against manufacturing constraints that are beyond the scope of this paper.

\begin{figure}[!ht]
\centering
  \includegraphics[width=0.75\linewidth]{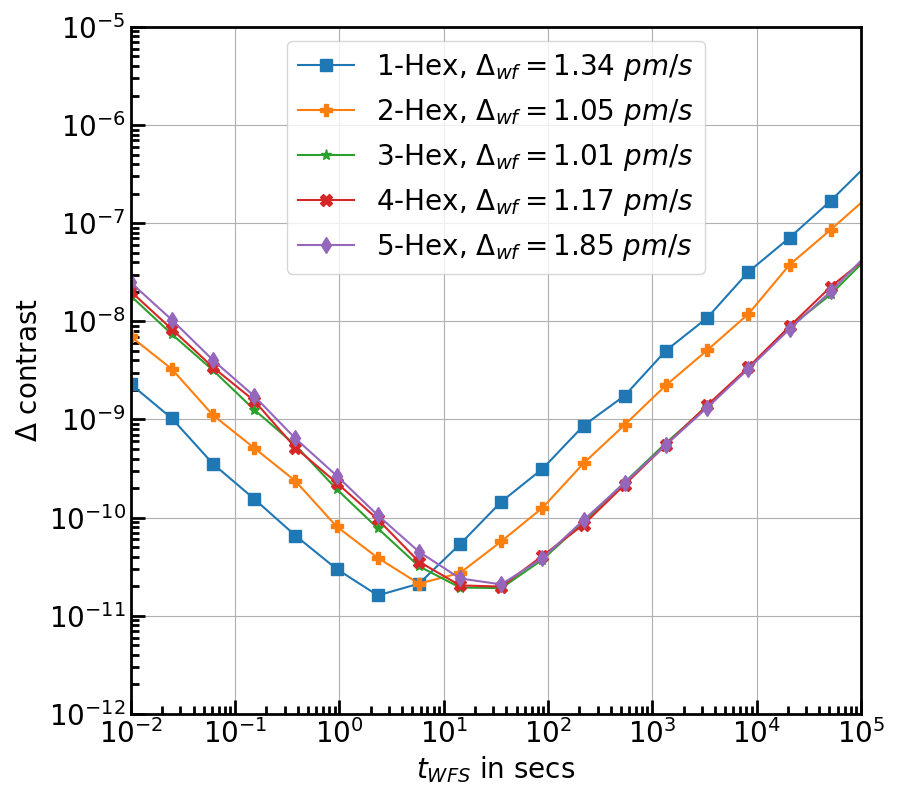}
  \caption{Close-loop contrast stability at different wavefront-sensor exposures for all SCDA designs along with their optimal wavefront-error temporal tolerances. We see a general trend that with an increase in segment numbers, the requirements become less stringent, and the 5 Hex design can withstand higher wavefront error drift to achieve the target contrast.} \label{fig:temporal_designs}
\end{figure}

\begin{figure}
\centering
 Contrast Error Budget for the 5-Hex Design Considering Wavefront Drift 
     \begin{subfigure}{0.49\textwidth}
         \centering
         \includegraphics[page=1, height = 0.38\textheight, width=\linewidth]{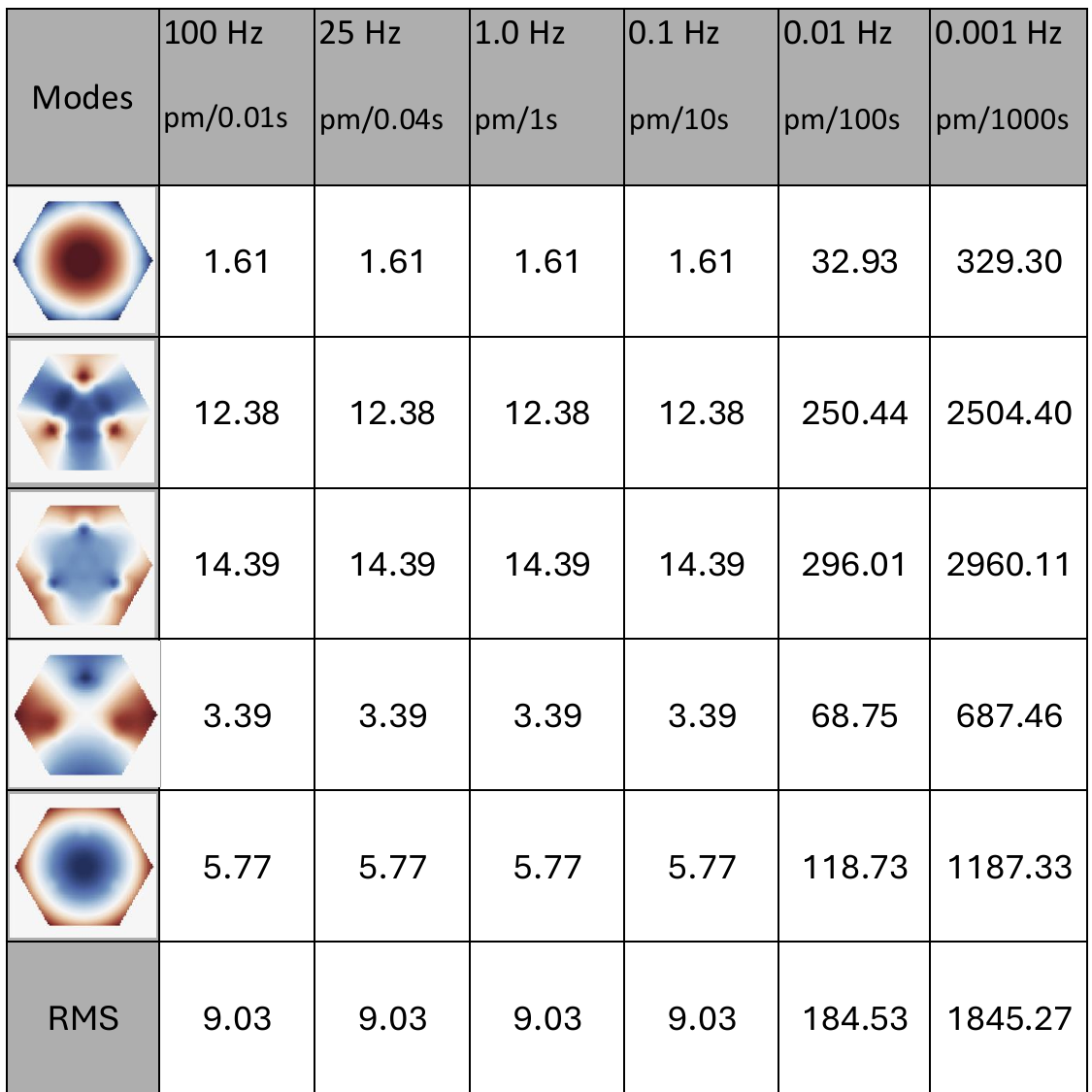}
         \caption{Tolerances}
         \label{fig:1-Hex temporal}
     \end{subfigure}
     \begin{subfigure}{0.49\textwidth}
         \centering
         \includegraphics[page=2, height = 0.38\textheight, width=\linewidth]{full_error_budget_5Hex.pdf}
         \caption{Contrast Allocation}
         \label{fig:2-Hex temporal}
     \end{subfigure}
    \caption{\textit{Left:} RMS dynamic wavefront error allocation across all segments required to achieve a DH contrast stability of $10^{-11}$. \textit{Right:} Corresponding mean contrast allocation for each of the modes in the DH.} 
    \label{fig:full_error_budget_5hex} 
\end{figure}

The optimal drift scales and wavefront-sensing exposure times obtained from the minima of curves in Fig.~\ref{fig:SCDA temporal} are used to compute mean temporal drifts across all segments for all of the SCDA designs. For a comparative analysis between different types of telescope geometries, we summarize the optimal mean close-loop WFE drifts(across five thermal modes, across all segments) in Fig.~\ref{fig:temporal_designs}. Our analysis estimates that the 3-Hex, 4-Hex and 5-Hex designs can achieve the desired $10^{-11}$ DH contrast in similar WFS time scales, with tolerance requirements decreasing progressively from 3-Hex to 4-Hex to 5-Hex. For the 1-Hex and 2-Hex geometries, we observe a trade-off between optimal WFS time and drift requirements. The optimal WFS time for 1-Hex design is smaller than that for the 2-Hex design, whereas the tolerance requirements are more stringent for the 2-Hex design. Overall, we observe that with an increase in segment numbers (or decrease in segment sizes), the mean-tolerance requirement decreases. 
For the 5-Hex design, we construct a contrast-error budget table (shown in Fig.~\ref{fig:full_error_budget_5hex}) to illustrate how the optimal drifts, WFS times obtained above can be utilized. We obtained an optimal WFS time of ${\sim}40$ s which corresponds to 0.02 Hz, and we use this information to populate the left hand side (below 0.01 Hz) of the error budget table via the PASTIS analysis, for the right-hand side (above 0.01 Hz) of the error budget table, we use the optimal scales obtained by the batch estimation algorithm. Note that, we used the exact same contrast allocation across all five thermal modes, as this is based on the PASTIS analysis, which allocates equal contrast per WFE mode. Similar coronagraph error budget tables can be generated for the other SCDA telescope geometries. 
\newpage
\section{Summary and Conclusions}\label{sec:conclude}
In this article, we study the impact of thermally induced surface deformation of a segmented primary mirror on the coronagraphic dark hole contrast, and set thermal stability requirements crucial to enable Earth-like planet detections. Thermal gradients and mechanical instabilities in a space-based observatory can degrade the coronagraphic dark-hole contrast, and limit the detection of planetary signals. Quantifying stability requirements for observatories in this scenario requires performing MC simulations with integrated finite element models of thermal, structural and optical behavior, and is a complex task to implement. Here, we use an alternative methodology that combines a matrix based approach for optical propagation and an analytical model based on statistical estimation theory to construct a simplified contrast-error budget. This approach provides a good estimate in the scenario when the telescope operates in a steady-state linear regime, the electric field in the image plane is a linear function of small phase perturbation, and the contrast in the DH is limited by this quadratic phase error. If the contrast is dominated by higher-order terms, full end-to-end simulations such as those described in Nemati et al. (2023)\cite{nemati2023analytical}, Krist et al. (2023)\cite{krist2023end} are required. In this study, we present a relatively faster approach, which may be less precise compared to those full end-to-end simulations. We use a finite element model of the segment-deformation caused by thermal gradients, and propagate this model using the PASTIS framework to evaluate allowable segment-level surface deformations, subsequently evaluate temperature requirements for individual segments. In particular, we extend the PASTIS static tolerance analysis to arbitrary FEM modes. Using the PASTIS analysis, we find that the inner segments are to be constrained more tightly than the outer ones. The PASTIS tolerances are static in nature, and doesn't specify the timescales in which these tolerance maps are to be maintained. We translate the tolerance maps into the temporal domain using a batch estimation algorithm based on close-loop control. For a comparative analysis between segment sizes and numbers, we utilize multiple segmented primary mirror architectures obtained from a survey conducted by the SCDA team for a ${\sim}6$ m-aperture diameter. We set tolerances both in terms of surface deformation and temperature gradient for each of the SCDA telescope aperture designs necessary to achieve a spatial average dark hole contrast of $10^{-11}$. Consistent with the static tolerance analysis, we observe that the outer segments are also temporally more loosely constrained compared to the inner ones. The requirements on these individual segments are tied to the architectures of the downstream coronagraphic instrument. In our design cases, the outer segments that are partially or fully blocked by the pupil apodizer mask are less constrained than the inner ones, as they have less impact on the image plane contrast. Close-loop control relaxes wavefront error requirements for these modes. Our study indicates that when establishing stability requirements for imaging Earth-like planets, the architectures of both the primary mirror and the high contrast imaging instruments should be optimized together, rather than individually. 

\section*{Disclosures}
The authors declare no conflict of interest.

\section*{Data, Materials, and Code Availability} 
A python package, named ULTRA was developed to support our analysis, and is publicly accessible at https://github.com/spacetelescope/ULTRA. The package makes use of Astropy\cite{price2022astropy}, Matplotlib\cite{hunter2007matplotlib}, NumPy\cite{harris2020array}, SciPy\cite{virtanen2020scipy},  Pandas\cite{pandas_cite, mckinney2011pandas}, Exoscene\cite{exoscene_citation}, HCIPy\cite{por2018high} and  PASTIS\cite{pastis-v2.2.0} packages. The segment-level thermal deformations data utilized in this study were obtained from BAE Systems Inc. and L3Harris Technologies and are available with their permission. 

\acknowledgments 
We would like to thank the reviewers for their valuable comments and suggestions, which have improved the presentation of this work. This work was co-authored by employees of BAE Systems (formerly Ball Aerospace) and L3Harris Technologies as part of the Ultra-Stable Telescope Research and Analysis (ULTRA) Program under Contract No. 80MSFC20C0018 with the National Aeronautics and Space Administration (NASA) (PI: L. Coyle), and by STScI employees under corresponding subcontracts No. 18KMB00077 and No. 19KMB00102 with BAE Systems (PI: R. Soummer, Sci-PI: L. Pueyo). This work was supported in part by the NASA Grant No. 80NSSC19K0120 issued through the Strategic Astrophysics Technology/Technology Demonstration for Exoplanet Missions Program (SAT-TDEM; PI: R. Soummer) and in part funded by the STScI Director’s Discretionary Research Fund. I. Laginja acknowledges the support by a postdoctoral grant issued by the Centre National d'Études Spatiales (CNES) in France. A. Sahoo thanks Anand Sivaramakrishnan for helpful discussions on numerical simulation of optical wave propagation. This paper presents an extended version of the work previously published in a conference paper\cite{sahoo2022segment}. 
\end{spacing}

\bibliography{report} 
\bibliographystyle{spiejour} 

\end{document}